\documentclass[aps,prl,reprint,showpacs,showkeys,footinbib,superscriptaddress]{revtex4-1}
\usepackage{amsmath,amssymb,bm}
\usepackage{graphicx}
\usepackage{times,txfonts}
\usepackage{xcolor}
\usepackage{epstopdf}
\usepackage[colorlinks=true,citecolor=blue,bookmarks=false]{hyperref}
\usepackage{lipsum}
\usepackage[all]{hypcap}
\usepackage{CJK}
\usepackage{textcomp}
\usepackage{tikz}
\usepackage{lmodern}
\usepackage[T1]{fontenc}

\makeatletter
\renewcommand{\bibsection}{}%
\makeatother

\usepackage[version=4,textfontname=sffamily,mathfontname=mathsf]{mhchem}

\begin{document}
\begin{CJK*}{UTF8}{bsmi}

\title{Efficient Magnetic Spin-Filtering and Persistent Spin-Currents in\\ 
Lifshitz-Transitioned Altermagnets: A Route to Open-Orbit Spintronics}

\author{Wun-Hao Kang (康文豪)}
\affiliation{Department of Physics and Center of Quantum Frontiers of Research and Technology (QFort), National Cheng Kung University, Tainan 70101, Taiwan}

\author{Yu-Ting Hsiao (蕭妤庭)}
\affiliation{Department of Physics and Center of Quantum Frontiers of Research and Technology (QFort), National Cheng Kung University, Tainan 70101, Taiwan}

\author{Jozef Genzor}
\affiliation{Institute of Physics, Slovak Academy of Sciences, 84511 Bratislava, Slovakia}

\author{Yaroslav Zhumagulov}
\affiliation{Institute of Physics,
Ecole Polytechnique F\'{e}d\'{e}rale de Lausanne (EPFL), CH-1015 Lausanne, Switzerland}

\author{Ming-Hao Liu (劉明豪)}
\affiliation{Department of Physics and Center of Quantum Frontiers of Research and Technology (QFort), National Cheng Kung University, Tainan 70101, Taiwan}

\author{Denis Kochan}
\affiliation{Department of Physics and Center of Quantum Frontiers of Research and Technology (QFort), National Cheng Kung University, Tainan 70101, Taiwan}
\affiliation{Institute of Physics, Slovak Academy of Sciences, 84511 Bratislava, Slovakia}

\begin{abstract}
Altermagnets offer a unique venue for spin transport due to their vanishing net magnetization and momentum-dependent spin splitting. We demonstrate that a homotopic Lifshitz transition in two-dimensional altermagnets creates a regime where carriers are confined to geometrically protected, spin-selective open channels. These channels originate from non-contractible Fermi contours and act as metallic analogues of topological edge modes: they are sharply directional, spin-pure, and protected by Fermi-surface winding rather than an energy gap or boundary confinement. We predict three striking magneto-transport signatures of such topologically reconfigured altermagnets: open-orbit focusing with perfect lensing and retroreflection, high-efficiency magnetic spin filtering, and chirality-tunable spin persistent currents in altermagnetic nanotubes. Our results establish altermagnets as a platform where Fermi-surface winding directly engineers spin transport, bypassing the requirements for ferromagnetism or strong spin-orbit coupling. These findings identify Lifshitz-transitioned altermagnets as a route to topology-enabled spintronics that transcends the limitations of conventional edge-state paradigms.
\end{abstract}
\maketitle
\end{CJK*}

The search for robust spin transport mechanisms that do not rely on ferromagnetic order is a central theme of modern spintronics \cite{Zutic2004Spintronics,Sinova2015SpinHall,Baltz2018AFMSpintronics,DalDin2024AFMSpintronicsBeyond,Manchon:RevModPhys.91.035004}. In most established platforms, spin control is achieved either by exchange fields in ferromagnets or by spin--orbit coupling in nonmagnetic materials. Altermagnets (AMs)~\cite{Smejkal2022Beyond,PRXEdditor2022,Libor2022Emerging,Song2025FunctionalMaterials,Jungwirth2026:Nature} open a qualitatively different route. They host spin-split electronic bands despite vanishing net magnetization, and thereby combine magnetic functionality with the absence of macroscopic stray fields. This makes them promising candidates for spin-current generation and spin filtering \cite{Gonzalez-Hernandez-PRL2021,Smejkal2022GMRTMR,Bose2022Tilted,Bai2022Observation,Zexin2022Anomalous,Samanta2025SpinFiltering,Fu2025AllElectrical,Bai2024,FenderJACS2025,Tamang:magnetism5030017,DanruQu:PhysRevLett.133.056701}.

Most current discussions of altermagnetic transport focus on the existence of spin splitting itself. A less explored question is whether the topology of the spin-resolved Fermi contour can by itself generate qualitatively new spin-transport regimes. This question is especially natural in two dimensions, where a Fermi contour is a closed curve on the Brillouin-zone torus and can therefore undergo changes not only of shape but also of homotopy class \cite{Lifshitz1960,Blanter1994ETTReview}. Across a Lifshitz transition, spin-split Fermi contours of AM may evolve from contractible closed loops to non-contractible open lines that wind around the torus. Such a transition does not merely reshape the Fermi surface: it reorganizes the available carrier trajectories and can fundamentally alter transport \cite{Lifshitz1957,Lifshitz1960,Kaganov_Lifshitz_1979,Blanter1994ETTReview}.

Here we show that this mechanism is realized in a two-dimensional $d$-wave AM and leads to a distinct open-orbit phase with direct spintronic consequences. In the open-orbit regime the corresponding quasiparticles propagate through effectively one-dimensional bulk channels. Remarkably, the two spin species are funneled into perpendicular directions, so that the metallic bulk itself acts as a spin-selective network of 
one-dimensional (1D) transport paths. In this sense, spin-momentum locked open-orbit phase provides a bulk analogue of directional topological transport: its key organizing principle is not a boundary state or a band Chern number, but the winding of the spin-resolved Fermi contour on the Brillouin-zone torus. 

This simple observation has several striking consequences. First, in a perpendicular magnetic field the open channels support transverse magnetic focusing (TMF) with perfect retroreflection when the focal points coincide with the sample edge. Second, in multiterminal Hall-bar geometries they enable strong spin filtering and nonlocal spin-current generation. Third, when the same system is wrapped into a nanotube, the interplay of open-orbit transport and quantized circumferential motion yields chirality-dependent persistent currents that are either fully spin-polarized or entirely quenched. Together, these results identify Lifshitz-transitioned AMs as a platform where Fermi-surface topology becomes a practical tool for engineering spin transport. Remarkably, such non\-tri\-vially winding Fermi contours---and hence the regime of \textit{open-orbit AM spintronics} are genuine only for $d$-wave AMs and being topologically excluded for $g$-, $i$-, or higher-wave AM classes.

\begin{figure*}[th]
    \centering
    \includegraphics[width=2\columnwidth]{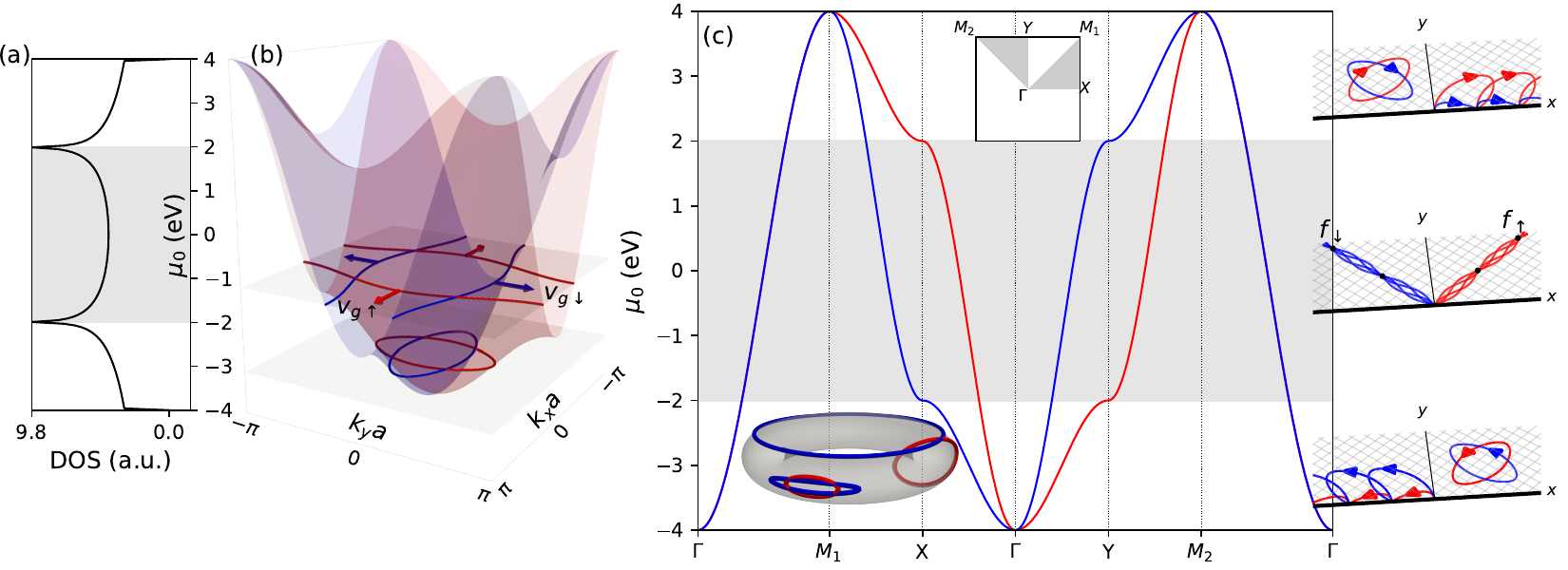}
    \caption{Spectral and homotopic characteristics of 2D $d$-wave AM. 
    Doping dependence of (a)~density of states (DOS) per unit area; (b)~Fermi contour topologies; and (c)~electronic band structure along the high-symmetry path. All based on the model Hamiltonian $\mathcal{H}$, Eq.~(\ref{eq:hamiltonian}), placed on the square lattice with the primitive vectors 
    $\mathbf{a}_{1}=(a,0)$ and $\mathbf{a}_{2}=(0,a)$. The electronic states are 
    spin-polarized, in all panels---the up and down spin states are colored by red and blue. Before the first, $\mu_0 < -2=-4\,\text{min}(|t_0|,|t_j|)$\,eV, and after the second, $\mu_0 > +2=4\,\text{min}(|t_0|,|t_j|)$\,eV, van Hove singularity the Fermi contours are homotopically trivial (contractible to a point). In between, i.e.~for 
    $\mu_0\in(-2,+2)$\,eV---shaded areas---the Fermi contours are homotopically nontrivial and opposite spin polarizations propagate with perpendicular group velocities $\mathbf{v}_{g\uparrow}$ and $\mathbf{v}_{g\downarrow}$. Corresponding homotopic visualization of the open and closed orbits on the 2D torus is displayed in the lower inset of panel (c). 
    The upper inset in panel (c) shows the $\mathbf{k}$-path in the 1st Brillouin zone along which the electronic band structure is calculated. The three insets on the right of panel (c) show the semi-classical AM cyclotron orbits at $\mu_{0}=-3$, $0$, and $3$~eV in the half-plane $y>0$ at $B_z=6.3$~T. Correspondingly from the bottom to top: electron-like, $\text{sgn}(1/m^*)=(+,+)$, closed and skipping orbits; 
    open-orbits with a hyperbolic mass tensor $\text{sgn}(1/m^*)=(+,-)$; and the hole-like, $\text{sgn}(1/m^*)=(-,-)$, closed and skipping orbits. 
    The middle inset reveals the \emph{open-orbit magnetic focusing}: a sequence of magnetic-field-defined focal hotspots, $\mathrm{f}_{\uparrow}$ and $\mathrm{f}_{\downarrow}$, generated by the hyperbolic mass tensor, at which spin-polarized trajectories converge independently of their initial injection velocity, for animation see SM~\cite{supp}.}
    \label{fig:Fig1_main_torus}
\end{figure*}

While metallic AMs such as CrSb~\cite{Zheng:AdvancedScience2024,Yang2025CrSb,Ding2024Large} and possibly also RuO$_2$~\cite{Fedchenko2024RuO2TRSB,HUSSAIN2025417723,choi:NanoConvergence2026,DanruQu:NanoLett2026,DanruQu:PhysRevLett.133.056701} provide key experimental benchmarks, their high carrier densities make spin-resolved open Fermi contours difficult to tune.
Rather than ruling out open-orbit-driven spin physics in AM, the way forward lies in more reconfigurable AM platforms, such as van der Waals materials \cite{Regmi2025CoNb4Se8,Liu:PhysRevLett.133.206702,Gonzalez2025}, semiconductors \cite{Amin:Nature2024,Betancourt2023MnTeAHE,Belashchenko:PhysRevLett.134.086701}, oxide/perovskite families \cite{Naka2025AltermagneticPerovskites,Streltsov2025Altermagnetism6H}, ultracold-atom AM simulators \cite{Das2024Realizing}, and multiferroic settings \cite{Gu2025FerroelectricSwitchable,Duan2025AntiferroelectricAltermagnets,Peng2026FerroelasticAltermagnetism,Camerano2025}, where doping, strain, confinement, interfaces, twisting and electric fields offer direct control over Fermi-surface topology.  
Motivated by this prospect, in this Letter we investigate previously unexplored topological properties of two-dimensional (2D) AMs, focusing on homotopic transitions, efficient magnetic spin-filtering, TMF and spin persistent currents. We use the minimal two-band tight-binding Hamiltonian of $d$-wave AM  \cite{Libor2022Emerging,Giil2024Superconductor,YutaroPRB2024Spinpolarized,Maier2023Weak,Brekke2023TwoDimensional,Roig2024Minimal}:
\begin{align}
\mathcal{H} &= -\mu_{0}\sum_{m}\sum_{\sigma}c^{\dagger}_{m\sigma}\left[s_{0}\right]_{\sigma\sigma}c^{\phantom{\dagger}}_{m\sigma}\notag\\
&+\sum_{\langle m,n\rangle}\sum_{\sigma}c^{\dagger}_{n\sigma}\left(t_{0}\left[s_{0}\right]_{\sigma\sigma}
+\nu_{n\leftarrow m}\,t_{j}\left[s_{z}\right]_{\sigma\sigma}\right)c^{\phantom{\dagger}}_{m\sigma}
\label{eq:hamiltonian}
\end{align}
that resides on a square lattice given by primitive lattice vectors
$\mathbf{a}_{1}$ and $\mathbf{a}_{2}$. 
Here $c^{(\dagger)}_{m\sigma}$ are annihilation (creation) operators acting on electrons with spin $\sigma$ on lattice site $m$, and $s_{0/x/y/z}$ are, correspondingly, 2-by-2 identity and Pauli spin matrices. The first term in $\mathcal{H}$ tunes a number of particles via the chemical potential $\mu_{0}$~\footnote{For this reason it will be more appropriate to call $\mathcal{H}$ as the grand-canonical or grand-Hamiltonian, however, in order to simplify terminology we use just a plain term ``Hamiltonian''.}. The second term represents nearest neighbor hybridization comprising of an isotropic, spin-unresolved hopping $t_{0}$, and of an anisotropic, spin-selective hopping $t_{j}$ imparting the characteristic $d$-wave spin texture of AM; 
in the main text we employ $t_0=-1$\,eV and $t_j=-0.5$\,eV, while Supplemental Material (SM)~\cite{supp} discusses other parameter regimes. 
Hopping sign factor $\nu_{n\leftarrow m}$ encodes a direction of the nearest neighbor hopping between site $m$ and $n$. It is equal to $+1$ ($-1$) if the hopping path ${n\leftarrow m}$ is aligned along $\pm\mathbf{a}_1$ ($\pm\mathbf{a}_2$) lattice vectors. 
The spin-projection $\sigma$ is a good quantum number, and the corresponding eigenspectrum of $\mathcal{H}$ reads 
\begin{equation}\label{eq:EigenSpectrum}
\begin{aligned}
E_\sigma(\mathbf{k},\mu_0)
=
&2t_{+,\sigma}\cos{(\mathbf{k}\cdot \mathbf{a}_1)}+2t_{-,\sigma}\cos{(\mathbf{k}\cdot \mathbf{a}_2)}-\mu_0\,,
\end{aligned}
\end{equation}
where $t_{\pm,\sigma}=t_0\pm\sigma t_j$.
Figures~\ref{fig:Fig1_main_torus}(a) and (b) show, respectively, the density of states (DOS) and the evolution of the Fermi contours, $E_\sigma(\mathbf{k},\mu_0)=0$, revealing clear signatures of the van Hove singularities and homotopic transitions.

\medskip
\paragraph{\textbf{Homotopic Lifshitz transitions (HLTs) and spin magneto-filtering.}} 
HLT \cite{Lifshitz1957,Lifshitz1960,Kaganov_Lifshitz_1979} is a spectral crossover when the Fermi surface $\mathcal{S}_{\text{F}}$ changes its homotopy group. 
Although HLT is accompanied by the van Hove singularity it does not, in general, entail a gap closing and reopening, unlike topological transitions driven by changes of Chern numbers.
Conventional Lifshitz transitions, akin to HLTs, normally refer to transitions that alter the number of connected components of $\mathcal{S}_{\text{F}}$ while leaving their homotopy class unchanged.
Rather than delving into the subtleties of such \textit{2.5-order phase transitions} in the classic sense of Lifshitz \cite{Lifshitz1960}---we instead focus on one particular HLT rendered 
by 2D AMs. 

As shown in Figs.~\ref{fig:Fig1_main_torus}(a) and (b), varying a doping in the AM band dispersions $E_\sigma(\mathbf{k},\mu_0)$, Eq.~(\ref{eq:EigenSpectrum}), gives rise at chemical potentials $\mu_0=\pm 4\,\text{min}(|t_0|,|t_j|)$ to pronounced van Hove singularities, and also to the spin-polarized closed, and open Fermi contours. 
The case of closed orbits, 
$\mu_0\in \boldsymbol{[}-4\,\text{max}(|t_0|,|t_j|),-4\,\text{min}(|t_0|,|t_j|)\boldsymbol{)}\,\bigcup\,$
$\boldsymbol{(}4\,\text{min}(|t_0|,|t_j|),4\,\text{max}(|t_0|,|t_j|)\boldsymbol{]}$---independently of their spin-po\-la\-ri\-za\-tion---is homotopically trivial, as these contours can be continuously contracted to a point, while the case of open spin-split Fermi contours for $\mu_0\in\boldsymbol{(}-4\,\text{min}(|t_0|,|t_j|),4\,\text{min}(|t_0|,|t_j|)\boldsymbol{)}$, is homotopically non-trivial. 
The fundamental group of the underlying Brillouin zone torus $T^2$ equals $\pi_1(T^2)=\mathbb{Z}_{k_x}\times\mathbb{Z}_{k_y}$. This means that any loop in the Brillouin zone is characterized by a pair of integers---\emph{winding vector of the Fermi contour on the Brillouin torus},
\begin{equation}
\mathbf{w} = (w_{k_x}, w_{k_y}) \in \mathbb{Z}\times\mathbb{Z},
\end{equation}
which counts how many times the contour wraps around the two fundamental cycles of $T^2$.
The different regimes are then naturally classified; $\mathbf{w} = (0,0)$ for closed (contractible) orbits, and $\mathbf{w} \neq (0,0)$ for the open Fermi contours. 
Inspecting Fig.~\ref{fig:Fig1_main_torus}(b), the open spin-up Fermi contours (red lines) acquire winding vector $\mathbf{w}_\uparrow=(0,\pm 1)$, whereas the spin-down ones (blue lines) carry $\mathbf{w}_\downarrow=(\pm 1,0)$. Hamiltonians of $d$-wave AMs with \textit{higher winding vectors} that directly generalizes Eq.~(\ref{eq:EigenSpectrum}) 
are further scrutinized in~\footnote{Obviously, Hamiltonian $\mathcal{H}=\sum_{\mathbf{k}\sigma}\,E_\sigma(\mathbf{k})\, c^\dagger_{\mathbf{k}\sigma}c^{\phantom{\dagger}}_{\mathbf{k}\sigma}$ with energy dispersion
\[
\begin{aligned}
E_\uparrow(\mathbf{k})
=
-\mu_0 &+ 2t_{1}[\cos{(\mathbf{k}\cdot \mathbf{a}_1)}+\cos{(\mathbf{k}\cdot \mathbf{a}_2)}]\\
&+
2t_{2}\cos{(m \mathbf{k}\cdot \mathbf{a}_1 - n \mathbf{k}\cdot \mathbf{a}_2)}
\end{aligned}
\]
and 
$E_{\downarrow}(\mathbf{k}) = E_\uparrow(\text{Rot}_{\pm 90^\circ}\mathbf{k})$, gives 
rise for different $\mu_0$ and $t_2$ to a family of open Fermi contours with higher winding vectors 
$\mathbf{w}_\uparrow=(\pm\tilde{m},\pm\tilde{n})$ 
and 
$\mathbf{w}_\downarrow=(\pm\tilde{n},\pm\tilde{m})$, where $\tilde{m}=0,\dots m$ and $\tilde{n}=0,1,\dots,n$. 
Special case, $m=1,n=0$, corresponds to Eq.~(\ref{eq:EigenSpectrum}) upon the identification:
$t_{+,\uparrow}=t_1+t_2=t_{-,\downarrow}$, and 
$t_{-,\uparrow}=t_1=t_{+,\downarrow}$. To see the appearance of higher winding vectors, one typically considers a large magnitude of $t_2$, then the Fermi contours are specified
by lines $m \mathbf{k}\cdot \mathbf{a}_1-n \mathbf{k}\cdot \mathbf{a}_2=0\,(\text{mod}\,\pi)$. For a square lattice with $\mathbf{a}_1=(1,0)$ and 
$\mathbf{a}_2=(0,1)$, this gives rise to a family of 2D-torus curves $k_x/k_y=n/m$ for spin up states, and $k_x/k_y=m/n$ for the spin down ones.}. 

As $E_\uparrow(\mathbf{k},\mu_0) = E_\downarrow(\text{Rot}_{\pm 90^\circ}(\mathbf{k}),\mu_0)$, see Figs.~\ref{fig:Fig1_main_torus}(b,c) and Eq.~(\ref{eq:EigenSpectrum}), the open-orbit states with opposite spins propagate in $\mathbf{k}$-space along mutually perpendicular directions.
Consequently, under a homogeneous transverse magnetic field $\mathbf{B}=(0,0,B_z)$, carriers with opposite spins will reciprocally trace perpendicular real-space trajectories (rescaled by $\hbar/(e |B_z|)$ and rotated by $90^\circ$). 
This furnishes a ho\-mo\-to\-py-pro\-tec\-ted pathway toward \textit{high-effi\-cien\-cy mag\-neto-spin-filtering} and \textit{robust spatial spin-separation} in the Lifshitz-transitioned $d$-wave AMs. 

Moreover, whether the given magneto-trajectory remains closed or becomes open curve correlates with the signature of the inverse effective-mass tensor $(1/m_\sigma^*)^{ab}=\hbar^{-2}\partial^2_{k_ak_b}E_\sigma(\mathbf{k},\mu_0)$,
that changes its definiteness at van Hove singularities $\mu_0=\pm 4\,\text{min}(|t_0|,|t_j|)$.
Positive signature, \textbf{Region~I}, occurring for dopings $\mu_0\in\boldsymbol{[}-4\,\text{max}(|t_0|,|t_j|),-4\,\text{min}(|t_0|,|t_j|)\boldsymbol{)}$ corresponds to closed electron-like orbits, while negative signature, \textbf{Region~II}, realized for $\mu_0\in\boldsymbol{(}4\,\text{min}(|t_0|,|t_j|),4\,\text{max}(|t_0|,|t_j|)\boldsymbol{]}$ covers closed hole-like states. The most interesting hyperbolic (or saddle-point) signature, \textbf{Region~III}, furnishing open Fermi contours is present for $\mu_0\in\boldsymbol{(}-4\,\text{min}(|t_0|,|t_j|),4\,\text{min}(|t_0|,|t_j|)\boldsymbol{)}$. Representative solutions of the semi\-clas\-sical equations of motion, $\dot{\mathbf{r}}_\sigma=\mathbf{v}_\sigma=\tfrac{1}{\hbar}\partial_{\mathbf{k}}E_\sigma(\mathbf{k})$ and $\hbar\dot{\mathbf{k}}=-e(\mathbf{v}_\sigma\times\mathbf{B})$ are shown for both spin projections and for three representative dopings---signatures of $(1/m_\sigma^*)^{ab}$---in the rightmost insets 
of Figs.~\ref{fig:Fig1_main_torus}(c).

\medskip
\begin{figure}[h!]
    \centering
    \includegraphics[width=\linewidth]{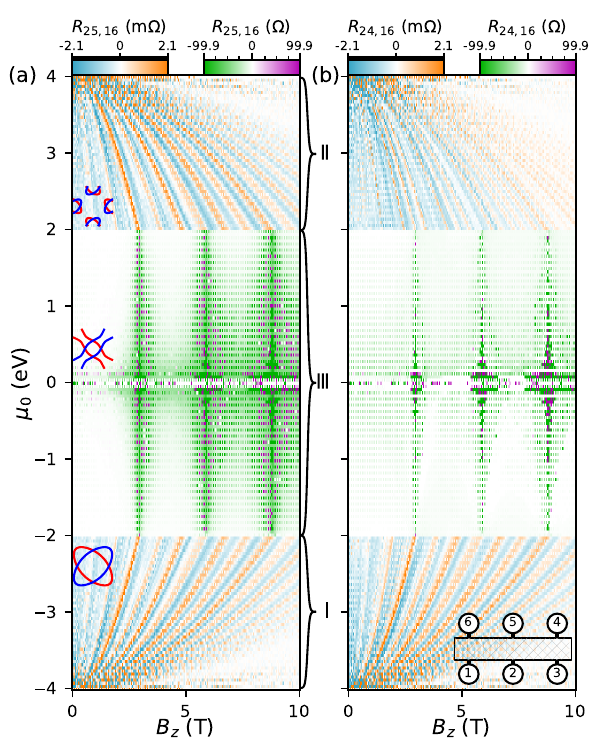}
    \caption{\textsc{kwant}-computed four-point terminal resistances $R_{25,16}$ and $R_{24,16}$, panels~(a)~and~(b), versus out-of-plane field $B_z$ and doping $\mu_0$. Simulated $6$-terminal Hall-bar [inset in (b)] consists of AM (at $45^{\circ}$ with respect to bar edges) and $6$ metallic leads, each with $130$-propagating transverse modes. 
    Color bars and ranges of $R$---$\rm m\Omega$~vs.~$\Omega$---are different for the regions supporting closed 
    (I and II) and open (III) Fermi contours. Insets in (a) show the representative topographies of spin-resolved (up/down in red/blue) Fermi contours at 
    $\mu_{0} = -3$~(I), $-1$~(III), $3$~(II)~eV.
    The non-local $R$ in closed-orbit Regions I and II show characteristic TMF oscillations due to electron-like (positively-defined $1/m^*$) and hole-like (negatively-defined $1/m^*$) skipping orbits.
    Contrary, in Region III (hyperbolic effective mass $1/m^*$), magneto-transport is effectively 1D---spin up (down) states propagate along directions given by the primitive lattice vectors---which yields spin-polarized current between injector \textcircled{\scriptsize{$2$}} and collector \textcircled{\scriptsize{$4$}} (or \textcircled{\scriptsize{$6$}}). This makes $I_{24}\ (I_{26})$ larger than $I_{25}$, and hence $R_{25,16}>R_{24,16}$ in Region III. Calculations are done at zero temperature, the absence of thermal smearing causes sharp transitions of $R$'s at the boundaries between Regions I and III and between III and II.}
    \label{fig:Fig2_main_map}
\end{figure}

\paragraph{\textbf{TMF and open-orbit magneto-focusing.}} To corroborate our semiclassical framework and homotopy-based analy\-sis, we perform numerical transport simulations using \textsc{kwant}~\cite{groth_kwant_2014} to calculate non-local resistances $R_{\alpha\beta,\gamma\delta}=V_{\gamma\delta}/I_{\alpha\beta}$ 
as functions of magnetic field and doping, see Fig.~\ref{fig:Fig2_main_map}.
We simulate six-terminal Hall-bar geometry shown in the inset of Fig.~\ref{fig:Fig2_main_map}(b), with semi-infinite metallic leads---each featuring parabolic band dispersion and $130$ transverse modes to ensure reliable transport calculations. The leads are $100\,\mathrm{nm}$ wide, displaced by $600\,\mathrm{nm}$ from the sample edges, and laterally separated by $2\,\mu\mathrm{m}$. The probe region has width $W=1\,\mu\mathrm{m}$, and the AM lattice constant is set to $a=1\,\mathrm{nm}$~\footnote{While this is about $2$--$3$ times larger than the lattice constant in typical metallic AMs, like MnSe ($3.67\,\mathrm{\AA}$), MnTe ($4.49\,\mathrm{\AA}$), and RuO$_2$ ($4.49\,\mathrm{\AA}$), it substantially lowers the computational cost without qualitatively affecting physics.}. Here we consider the particularly interesting case of an AM rotated by $45^\circ$ with respect to the Hall-bar edges; other orientations are presented in the SM~\cite{supp}.
Comparison of $R_{25,16}$ and $R_{24,16}$ for two different source-drain paths, $I_{2\rightarrow 5}$ and $I_{2\rightarrow 6}$, are shown in Figs.~\ref{fig:Fig2_main_map}(a) and (b). In the \textbf{Region~I} the two signals behave almost identically---closed electron-like orbits evolve into edge-propagating skipping orbits and generate conventional TMF oscillations~\cite{RevModPhys.71.1641}.  
In the closed-orbit \textbf{Region~II}, however, $R_{25,16}$ exhibits more pronounced oscillations than $R_{24,16}$. 
The perceptibly noisier TMF response of $R_{24,16}$ is due to more extended route from the drain~\textcircled{\scriptsize 4} to the voltage probe~\textcircled{\scriptsize 6}. 
Lorentz-force deflection toward probe~\textcircled{\scriptsize 6} forces hole-carriers into longer, bouncing-off-the-edge paths, which increases boundary scattering and defocusing and therefore impedes the TMF.
In the \textbf{Region~III}, $R_{25,16}$ is generally larger than $R_{24,16}$ due to a spatial alignment of current paths carried by open-orbits that are preferentially spreading in a direction from the source~\textcircled{\scriptsize 2} to drain~\textcircled{\scriptsize 4}, see the semiclassical simulation in the right insets of
Fig.~\ref{fig:Fig1_main_torus}(c). Hence the current entering the denominator of $R_{\alpha\beta,\gamma\delta}=V_{\gamma\delta}/I_{\alpha\beta}$ increases, which lowers the value of $R_{24,16}$. 

Generally, in the open-orbit range, the unique AM dispersion produces pronounced, doping-independent peaks in $R$ (green coniferous shapes in centers of Fig.~\ref{fig:Fig2_main_map}) at
\begin{equation}
B_{z,n} = \pm n\,\frac{\Phi_0}{\sqrt{2} W a},
\end{equation}
where $W$ is the width of the Hall-bar, $a$ lattice constant and $\Phi_0=h/e$ the flux quantum. 
A giant magnetoresistance due to open-orbits was reported in $\gamma$-PtBi$_2$~\cite{PhysRevResearch.2.022042}.
Semiclassically, $B_{z,n}$ here corresponds to \emph{magnetic focusing conditions for the effectively 1D open-orbits}, see Fig.~\ref{fig:Fig1_main_torus}(c);
electrons injected from terminal~\textcircled{\scriptsize 2} converge onto a series of spin-polarized hotspots, $\mathrm{f}_{\uparrow}$ and $\mathrm{f}_{\downarrow}$. When any of these focal points align with the opposite edge of the Hall bar, the carriers undergo perfect retro-reflection (contra-current), which generates a singular response in the resistance $R$, for animation see SM~\cite{supp}. Conceptually, \textit{open-orbit magnetic focusing} establishes \textit{Lifshitz-transitioned AM} as a promising platform for \textit{spin-resolved electron-optic lensing, beam steering, and collimation}. 
Although our semiclassical picture is used only as a heuristic guide, the results in Fig.~\ref{fig:Fig2_main_map} are fully quantum coherent and thus include magnetic breakdown, anomalous broadening, coherent orbit-network formation~\cite{PhysRevB.104.035419,PhysRevB.109.L081103} and possibly also other saddle-point magneto-anomalies including, for example, the solid-state analogue of the Hawking-Unruh effect~\cite{PhysRevLett.123.156802,SUBRAMANYAN2021168470}. 

\medskip
\begin{figure}[th!]
\centering
\includegraphics[width=1\columnwidth]{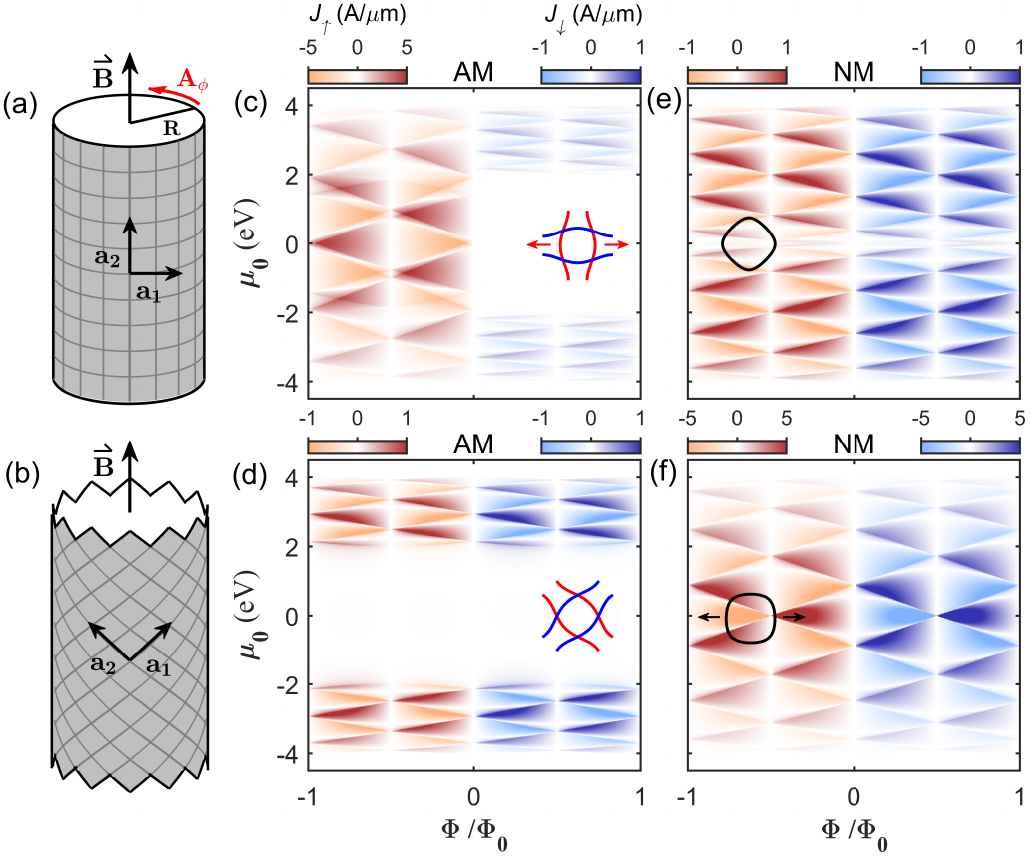}    
\caption{
Persistent spin currents (per unit length) $J_\uparrow$ and $J_\downarrow$ versus doping $\mu_0$ and magnetic flux $\Phi/\Phi_0$, for AM (central panels) and NM (right panels) nanotubes with the square (upper panels) and rhombic (lower panels) chiralities. Figures~(a)~and~(b)~show schematics of the nanotube geometries with respect to the axial magnetic field $\mathbf{B}$ and its polar gauge potential $\mathbf{A}_\phi$. Panels~(c,d) and (e,f) show, comparatively, $J_\uparrow(\Phi,\mu_{0})$ and $J_\downarrow(\Phi,\mu_{0})$ for ballistic (disorder-free) AM and NM nanotubes in weak magnetic fields employing $t_{0}=-1$\,eV, and correspondingly $t_{j}=-0.5$\,eV (AM) and $t_{j}=0$\,eV (NM). For AM tube with square-chirality, panel~(c);
tuning $\mu_{0}$ across the open-orbit range 
$\mu_{0}\in(-2,2)$\,eV makes PC perfectly spin-polarized rendering such nanotubes as \textit{topology-empowered spin-current generators}. In a sharp contrast, the AM nanotubes with rhombic-chirality, panel~(d), exhibit a complete quench of spin PCs within the same doping range,
making them ideal \textit{persistent-current insulators}.
For NM nanotubes with square-chirality, panel~(e), the spin currents are moderately suppressed near $\mu_0\approx 0$ (this suppression further enlarges under disorder; see SM), whereas the rhombic tube, panel~(f), remains persistent, however, both spin currents mix and produce no net spin polarization.
}

\label{fig:Altermagnet}
\end{figure}

\paragraph{\textbf{Persistent~spin~currents.}} Rolling 2D~AM in\-to a na\-no\-tu\-be---hollow thin-wall cylinder of radius $R$---and threading it with a homogeneous axial magnetic field $B$ creates another topologically non-trivial system. 
This geometry intertwines momentum-space homotopy (HLT tuned by the chemical potential $\mu_0$), the $U(1)$~holonomy associated with the Aharonov-Bohm (AB) phase set by the magnetic flux $\Phi=\pi R^2 B$, and the non-simply connected topology of the nanotube itself, whose fundamental group is $\pi_1(\text{nanotube})=\mathbb{Z}$.

For simplicity, we first describe geometry shown in Fig.~\ref{fig:Altermagnet}(a), where translational symmetry is preserved along 
the in-plane lattice vector $\mathbf{a}_{2}=a(0,1)$. Correspondingly, the AM plane is axially wrapped into a cylinder of radius $R$ along $\mathbf{a}_{1}=a(1,0)$ such that points $(x, y)$ and $(x + 2\pi R, y)$ are identified, in order to match the lattice periodicity $2\pi R = Na$ with $N\in\mathbb{N}$.
The AB-flux renders all nearest neighbor hoppings in $\mathcal{H}$,
Eq.~(\ref{eq:hamiltonian}), that are directed along $\pm\mathbf{a}_1$ by an 
additional $U(1)$-phase, $\exp(\pm i\frac{ 2\pi}{N}\frac{\Phi}{\Phi_{0}})$. 
Thus $\mathcal{H}$ loses translational symmetry along $\mathbf{a}_1$ and becomes flux-periodic 
$\mathcal{H}(\Phi+\Phi_0)=\mathcal{H}(\Phi)$. The combined effect of the AB-phase and circumferential momentum quantization gives in the present case, Fig.~\ref{fig:Altermagnet}(a), 
an exact eigenspectrum, $E_\sigma(m,k_y,\Phi,\mu_0)$, that takes the form of Eq.~(\ref{eq:EigenSpectrum}) 
with a replacement $k_x\rightarrow k_x(m,\Phi)=\frac{2\pi}{Na}(m+\frac{\Phi}{\Phi_0})$, where 
$m=0,1,\dots,N-1$; for computational details see SM~\cite{supp}.

Although we described the simplest wrapping defined by the primitive lattice vector $\mathbf{a}_1$, the construction extends straightforwardly to AM nanotubes with an arbitrary chirality vectors $\boldsymbol{\chi}=\chi_1\mathbf{a}_1+\chi_2\mathbf{a}_2$; the two simplest chiralities---square,
$\boldsymbol{\chi}=\mathbf{a}_1$, and rhombic, $\boldsymbol{\chi}=\mathbf{a}_1+\mathbf{a}_2$---are displayed in Figs.~\ref{fig:Altermagnet}(a)~and~(b). 
To determine persistent current (PC)~\cite{Kulik1970,KULIK20001880} of a nanotube of chirality $\boldsymbol{\chi}$ we find eigenvalues, $E_\sigma(m,k,\Phi,\mu_0)$, of the corresponding Hamiltonian $\mathcal{H}_{\boldsymbol{\chi}}(\Phi,\mu_0)$ and compute the spin-resolved grand-potential (per unit length) $\Omega_\sigma(\Phi,\mu_0)$, the corresponding spin PC (per unit length) reads:
\begin{equation}
J_{\sigma}(\Phi,\mu_0) 
= 
-\frac{\partial }{\partial \Phi}\,\Omega_\sigma(\Phi,\mu_0)
\,.
\end{equation}
Figures~\ref{fig:Altermagnet}(c,d) and \ref{fig:Altermagnet}(e,f) display, respectively, 
flux and doping dependencies of the PCs, $J_{\uparrow}$ and $J_{\downarrow}$, for 
the $d$-wave AM nanotube and its normal-metal (NM, $t_j=0$) counterpart at zero temperature. The upper panel shows results for
the square and lower for the rhombic-chirality. Effects of hard-wall boundary conditions (relevant for short nanotubes) and disorder, including details of analytical and numerical calculations are discussed in the SM~\cite{supp}.

Three salient features emerge strikingly in Figs.~\ref{fig:Altermagnet}(c)-(f) as the doping is tuned across the homotopically non-trivial region of $\mu_{0}\in(-4\,\text{min}(|t_0|,|t_j|),4\,\text{min}(|t_0|,|t_j|))$.

\textbf{(1)}~For the AM nanotube of square-chi\-ra\-li\-ty, Figs.~\ref{fig:Altermagnet}(a,c), 
the PC becomes perfectly spin-up polarized. 
The mechanism is rooted in the Fermi-contour topography and the directions of group velocities 
as illustrated in the inset of panel~\ref{fig:Altermagnet}(c).
The spin-up states (red) move along the circumference---one branch clockwise and the other counterclockwise. When the nanotube is threaded by a magnetic field, the resulting AB flux minimally couples to the underlying momenta. This coupling accelerates the group velocity of one branch, 
$\mathbf{v}_g + \frac{e}{m^*}\mathbf{A}_\phi$, while impeding the other, $-\mathbf{v}_g + \frac{e}{m^*}\mathbf{A}_\phi$. The resulting velocity imbalance yields 
a non-zero, perfectly spin-polarized PC. The spin-down states propagate axially and as they do not close into loops they are effectively decoupled from the AB flux.
Conceptually, AB-flux driven \textit{Lifshitz-transitioned AMs} can therefore act as 
\textit{topology-enabled platforms} towards \textit{functional spintronic devices} supplying \textit{perfectly spin-polarized persistent currents}.
The similar shape of the Fermi contours also explains a generation of PCs for 
rhombic NM nanotube close to the half-filling $\mu_0\approx 0$, Fig.~\ref{fig:Altermagnet}(f), the resulting currents are, however, spin-unpolarized.

\textbf{(2)}~By contrast, AM nanotubes with rhom\-bic-chi\-ra\-li\-ty, Figs.~\ref{fig:Altermagnet}(b,d), exhibit \textit{complete suppression of both spin-resolved persistent currents} throughout the homotopically non-trivial region of $\mu_{0}\in(-4\,\text{min}(|t_0|,|t_j|),4\,\text{min}(|t_0|,|t_j|))$.
As shown in the inset of panel~(d), states of opposite spin polarizations propagate along 
tilted directions $\pm\mathbf{a}_1$ and $\pm\mathbf{a}_2$---their helical orbits never close into loops and therefore remain entirely immune to the AB flux. While PC is a hallmark of mesoscopic coherence, it often poses a formidable challenge to spin- and charge-based electronics, introducing unwanted dissipation, signal interference, and decoherence. In this context, the quenching of PCs in AM nanotubes with rhombic-chirality is a remarkable asset. 
The Lifshitz-transitioned bands enforce open, helical trajectories that preclude a coupling to AB flux. In this sense, the \textit{non-trivial AM homotopy} not only \textit{suppresses persistent mesoscopic responses}; it transforms their absence into a decisive functional advantage. A similar quenching occurs also for the NM nanotubes with square-chirality,
Fig.~\ref{fig:Altermagnet}(e), in the immediate vicinity of half filling, $\mu_{0}\approx 0$, where the Fermi contours evolve into perfectly tilted squares that yield helical current propagation without a coupling to the AB-flux. 

\textbf{(3)}~In Regions~I~and~II, $J_\uparrow$ and $J_\downarrow$ are more strongly mixed, and the overall PC profiles of AM and NM nanotubes become more similar.

\medskip

\paragraph{\textbf{Altermagnetic spin-instabilities driven by HLT.}}
Enhanced DOS near the van Hove singularities typically drives system instabilities and triggers potential phase transitions. To scrutinize how homotopic transitions in AMs corroborate with many-body correlations in longitudinal and transverse spin channels 
we investigate bare, static spin-susceptibilities, $\chi^0_{^{\parallel/\perp}}(\mathbf{q})$, at room temperature~\footnote{Static, bare spin-polarization function (here called susceptibility) reads
\[
\chi^0_{\sigma\sigma'}(\mathbf q)
=
\frac1N
\sum_{\mathbf k}
\frac{
f\!\left[E_{\sigma'}(\mathbf k,\mu_0)\right]
-
f\!\left[E_\sigma(\mathbf k+\mathbf q,\mu_0)\right]
}{
E_\sigma(\mathbf k+\mathbf q,\mu_0)-E_{\sigma'}(\mathbf k,\mu_0)
},
\] 
where \(E_{\sigma}(\mathbf k,\mu_0)\) are AM eigen-energies given by 
Eq.~(\ref{eq:EigenSpectrum}), and $f(E)$ the corresponding 
Fermi-Dirac weights at temperature $T$. Consequently, the longitudinal and transverse polarizations equal: 
$\chi^0_{\parallel}(\mathbf{q})
=
\tfrac{1}{2}(\chi^0_{\uparrow\uparrow}(\mathbf{q})+\chi^0_{\downarrow\downarrow}(\mathbf{q}))$ 
and 
$\chi^0_{\perp}(\mathbf{q})
=
\tfrac{1}{2}(\chi^0_{\uparrow\downarrow}(\mathbf{q})+\chi^0_{\downarrow\uparrow}(\mathbf{q}))$.}.

\begin{figure}[th!]
\centering
\includegraphics[width=0.99\columnwidth]{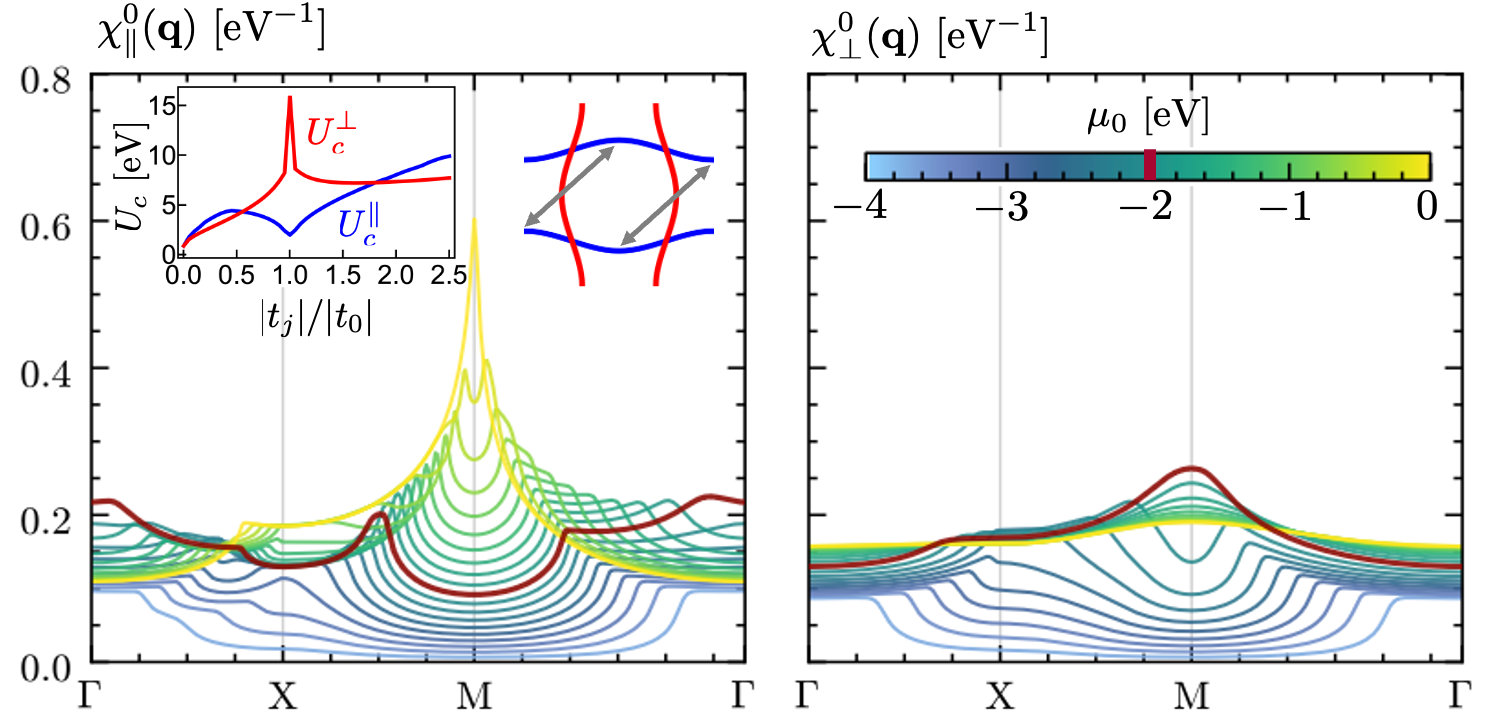}    
\caption{Longitudinal and transverse bare susceptibilities, $\chi^0_{^{\parallel/\perp}}(\mathbf{q})$, versus momenta and doping; calculations are for $T=300$\,K and $|t_j|/|t_0|=0.5$. At HLT, $\mu_0=-2$\,eV, brick-red lines,
$\chi^0_{^{\perp}}>\chi^0_{^{\parallel}}$, indicating ``paramagnon-like'' instability. Insets show: 
1) evolution of the critical interaction strength $U^{_{\parallel/\perp}}_{c}\equiv 1/\text{max}_{\mathbf{q}}[\chi^0_{^{\parallel/\perp}}(\mathbf{q})]$ at HLT for different $|t_j|/|t_0|$ and corresponding $\mu_0=-4\,\text{min}(|t_0|,|t_j|)$, and, 2)~Fermi contours at the half filling with M-point $(\pi,\pi)$ nesting. 
}
\label{fig:Suscept}
\end{figure}

Figure~\ref{fig:Suscept} illustrates the momentum-dependent bare susceptibilities \(\chi^0_{^{\parallel/\perp}}(\mathbf{q})\) for a representative set of chemical potentials. We vary \(\mu_{0}\) from \(-4\,\mathrm{eV}\), corresponding to zero filling, to \(\mu_{0}=0\), corresponding to half filling~\footnote{Due to particle--hole symmetry, \(\mu_{0}\mapsto -\mu_{0}\) the susceptibility profiles $\chi^0_{^{\parallel/\perp}}(\mathbf{q})$ at positive \(\mu_{0}\) are identical to those at the corresponding negative \(\mu_{0}\).}. At half filling the open Fermi contours become perfectly nested, see left inset in Fig.~\ref{fig:Suscept}, correspondingly the maxima of both $\chi^0_{^{\parallel/\perp}}(\mathbf{q})$ develop at $\mathbf q_c^{_{^{\parallel/\perp}}}=(\pm\pi,\pm\pi)$--M point. 
Moreover, in the most doping ranges, \(\chi^0_{^{\parallel}}>\chi^0_{^{\perp}}\), implying a softer longitudinal channel with spatially modulated magnitude of $\langle s_z(\mathbf{r})\rangle$ and zero $\langle s_{x/y}(\mathbf{r})\rangle$. Remarkably and uniquely to AMs, this trend reverses near the HLT, see the brick red lines in Fig.~\ref{fig:Suscept} corresponding to the van Hove singularity at \(\mu_0=-2\)\,eV. Around this doping the transverse spin-fluctuations softens---the global transverse maximum, \(\chi^0_{^{\perp}}(\text{M-point})\), is larger than the corresponding longitudinal one, \(\chi^0_{^{\parallel}}(\mathbf{q}\approx\text{$\Gamma$-point})\). 
This crossover points to an emergence of ``paramagnon-like'' spin-density-wave \cite{doniach1998greens} with
$\mathbf q_c^{\perp}\approx(\pm\pi,\pm\pi)$ owing to a spiral or helical in-plane spin texture, 
$\langle s_{x/y}(\mathbf{r})\rangle\neq 0$. This paramagnon-like instability shall emerge in Lifshitz transitioned AMs for $|t_j|/|t_0|\lesssim 0.55$ or $\gtrsim 1.8$, see inset in Fig.~\ref{fig:Suscept}. Importantly, the AM spin splitting already breaks the relevant spin symmetry at the single-particle level. The resulting paramagnon-like instability is therefore not a Goldstone mode, but a soft collective tendency in an explicitly spin-anisotropic background.

As our main objective is to establish \textit{open-orbit topological spintronics
in AMs}, we focus here on the stability of the open-orbit spin texture
against repulsive, spin-selective electron--electron interactions
$U_{\sigma\sigma'}(\mathbf q)$. A systematic analysis of the correlated phases
that may emerge in 2D AMs is left for future work. Within a Stoner-type criterion, the open-orbit state remains stable as long as the effective spin-selective interaction,
\(
U_{\sigma,-\sigma}(\mathbf q)-U_{\sigma,\sigma}(\mathbf q),
\)
does not exceed the critical interaction scale
\(
U_c^{_{\parallel/\perp}}
\equiv
1/\max_{\mathbf q}\chi^{0}_{_{\parallel/\perp}}(\mathbf q)
\)
in the corresponding longitudinal or transverse channel. In the present model,
the most vulnerable regime occurs near half filling, where the longitudinal
critical scale is only
\(
U_c^{\parallel}\simeq 1.6 |t_0| .
\)
By contrast, states close to the homotopic Lifshitz transition remain stable up
to substantially stronger interactions, with the transverse critical scale
approaching
\(
U_c^{\perp}\simeq 4 |t_0| .
\)

This indicates that the open-orbit regime is not merely a fine-tuned
single-particle feature, but survives over a sizable window of repulsive
spin-selective interactions.

\medskip
\paragraph{\textbf{Open orbits of g-wave AM on the double torus.}}

\begin{figure}[!htpb]
    \centering
    \includegraphics[width=\linewidth]{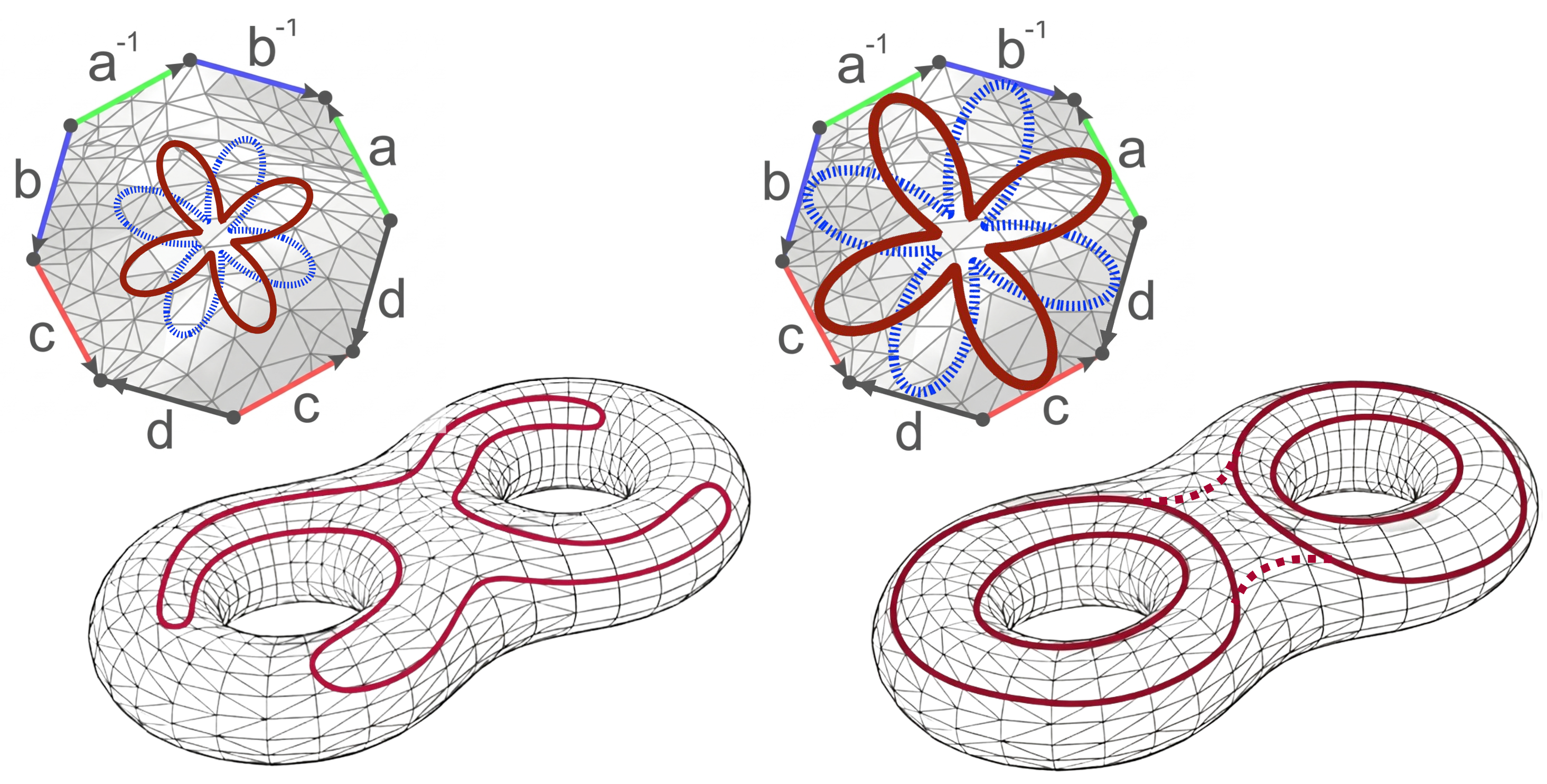}
    \caption{Evolution of loop topologies and homotopy classes with $g$-wave AM symmetry on the octagon fundamental domain and on the double torus $\Sigma_{g=2}$. The left panels show closed orbits, while the right panels illustrate the homotopically nontrivial case of open orbits at corresponding Lifshitz transition. The drawings on the double torus $\Sigma_{g=2}$ show only the spin-up states, represented by red curves. The corresponding spin-down states unfold analogously, but wind around the double torus in the complementary (through the holes) direction.}
    \label{fig:double-torus}
\end{figure}

A convenient way to understand why a $g$-wave AM does not generate
open Fermi orbits in the physical 2D Brillouin zone torus $T^2$, is to compare it 
with an auxiliary construction on a genus-two surface. The relevant octagonal fundamental polygon, when its edges are pairwise identified, defines an orientable Riemann surface $\Sigma_{g=2}$---double torus---rather than the Brillouin-torus. 
In this representation, the two four-petal Fermi contours with a $g$-wave AM symmetry 
are shown in Fig.~\ref{fig:double-torus}: the spin-up contour is plotted as a solid red
curve, while the spin-down contour is plotted as a dashed blue curve.

The left panel corresponds to a doping regime in which the Fermi contours remain
strictly inside the octagon and do not intersect its boundary. After the
octagon edges are glued, each four-petal contour therefore becomes a closed
contractible loop on $\Sigma_{g=2}$, one with a characteristic ``chromosome-X''
shape is shown explicitly. The right panel illustrates the Lifshitz point, where the Fermi contour
touches and coalesces with the octagon edges. Under the same edge
identifications, the red four-petal contour is then transformed into four
non-contractible cycles on $\Sigma_{g=2}$, each carrying unit winding around one
of the handles (fundamental cycles) of the double torus.

This construction shows that the apparent open-orbit structure of the
$g$-wave contour is naturally realized only after lifting the problem to the
genus-two surface $\Sigma_{g=2}$. Since the physical Brillouin zone remains genus-one
torus $T^2$, and not $\Sigma_{g=2}$, these homotopically non-trivial cycles on the
double torus do not correspond to genuine open Fermi orbits in the original
2D Brillouin zone.

The same topological reasoning extends to higher angular AM harmonics. For
example, an $i$-wave AM can be represented by an analogous construction
on a genus-three Riemann surface $\Sigma_{g=3}$, where its Fermi contours may
form non-contractible cycles on the auxiliary surface without implying open
orbits on the physical Brillouin-zone torus.

\medskip
\paragraph{\textbf{Summary and Outlook.}}
Spin-momentum locked Fermi-sur\-fa\-ce topology of $d_{x^2-y^2}$-al\-ter\-mag\-nets provides a theoretical route to engineered spin transport. Once the system is tuned across the homotopic Lifshitz transition, the electronic motion reorganizes into spin-selective open 1D channels owing to hyperbolic inverse mass tensor. In this homotopically-protected regime, the system supports a set of magneto-responses that are rarely combined within a single platform: magnetic spin filtering, nonlocal spin separation, perfect retroreflection and electron lensing, and chirality-controlled generation or quenching of fully spin-polarized persistent currents. The underlying functionalities are stable and resilient against electron-electron interactions, and are tied specifically to the $d$-wave spin-momentum locked altermagnets---excluding $g$- and $i$-wave classes. 

Beyond the minimal model studied here, this mechanism suggests several concrete opportunities. In planar devices, open-orbit altermagnets could operate as field-tunable spin routers, spin beam splitters, or nonlocal spin injectors without ferromagnetic contacts. In cylindrical or ring geometries, they offer a route to equilibrium spin-current sources and to mesoscopic architectures in which unwanted persistent charge currents are selectively suppressed by design. More broadly, the work highlights Lifshitz-transitioned altermagnets as a platform where topology of the Fermi surface, rather than topology of a filled band manifold and associated edge states, becomes the organizing principle for spintronic functionality. We predict that metallic room-temperature $d$-wave altermagnets KV$_2$Se$_2$O \cite{Jiang2025:NatPhys} and Rb${}_{1-\delta}$V$_2$Te$_2$O \cite{Zhang2025:NatPhys} are especially promising
platforms for \textit{open-orbit topological spintronics}---a path toward robust, symmetry-enabled spin control in systems that remain gapless, electrically accessible, and compatible with high-mobility transport.

\medskip

\makeatletter
\renewcommand{\bibsection}{}%
\makeatother
\paragraph{\textbf{References.}}
\bibliographystyle{apsrev4-2}
\bibliography{reference}


\medskip

\paragraph{\textbf{Methods.}}

\paragraph{Model and Fermi-contour analysis}

All calculations are based on the minimal spin-diagonal tight-binding model of a two-dimensional $d$-wave altermagnet introduced in Eq.~(1). Unless stated otherwise, we use $t_0=-1\,{\rm eV}$, $t_j=-0.5\,{\rm eV}$ and lattice constant $a=1\,{\rm nm}$. The spin-resolved band structures, density of states and Fermi contours were obtained by direct diagonalization of the Bloch Hamiltonian on a square-lattice Brillouin zone.

The topology of each spin-resolved Fermi contour was determined on the Brillouin-zone torus by solving equations $E_{\sigma}(\mathbf{k},\mu_0)=0$. Contractible contours were identified with trivial winding, while open contours in the repeated-zone representation were identified with non-contractible winding classes.

\paragraph{Semiclassical trajectories}

Semiclassical trajectories were used to visualize the real-space motion associated with the spin-resolved Fermi contours in a perpendicular magnetic field. The trajectories were obtained from the standard Bloch-electron equations of motion using a Runge--Kutta integration scheme. Particles were injected from the source contact with initial angles between $30^\circ$ and $150^\circ$ relative to the sample edge. At specular reflections we conserved energy, spin and canonical momentum parallel to the boundary. These calculations were used only as a semiclassical guide to the focusing mechanism; the transport results were obtained from fully quantum-coherent simulations.
\paragraph{Quantum-transport simulations}
Multi-terminal magnetotransport was calculated with the \textsc{kwant} package using the Landauer--B\"{u}ttiker formalism. The main simulations used a six-terminal Hall-bar geometry with the altermagnetic lattice rotated by $45^\circ$ relative to the device edges. The central region was described by Eq.~(\ref{eq:hamiltonian}), while the semi-infinite metallic leads were modeled by spin-degenerate parabolic bands. Each lead supported 130 propagating transverse modes. The lead width was $100\,{\rm nm}$, the separation between neighbouring leads was $2\,\mu{\rm m}$, the distance from the leads to the sample edge was $600\,{\rm nm}$, and the Hall-bar width was $W=1\,\mu{\rm m}$. 

A perpendicular magnetic field was introduced through Peierls phases in the hopping amplitudes. Non-local resistances were extracted from the calculated scattering matrix. All transport calculations shown in the main text were performed at zero temperature, which accounts for the sharp features at the boundaries between the closed- and open-orbit regimes. 
\setlength{\parskip}{0pt}
\paragraph{Nanotube persistent currents}
Persistent currents were calculated by rolling the 2D altermagnetic lattice into cylindrical geometries with different chiralities. The axial magnetic field was included through the corresponding Aharonov--Bohm flux, implemented as a Peierls phase along the circumferential direction. For each chirality and flux value we diagonalized the flux-dependent Hamiltonian and computed the spin-resolved grand potential at zero temperature. The persistent spin current was then obtained by differentiating the grand potential with respect to the flux. Analytical results, finite-length effects and disorder were analyzed separately in the Supplementary Material. Disorder was modeled by random on-site potentials drawn independently from a uniform distribution in the interval $[-0.1,0.1]\,{\rm eV}$, and disorder-averaged results were obtained from ten independent realizations.
\paragraph{Interaction stability}
The stability of the open-orbit spin texture against spin-selective repulsive interactions was analyzed within a Stoner-type criterion. Bare longitudinal and transverse spin susceptibilities were evaluated from the non-interacting band structure, on finite $\mathbf{k}$-grid ($480\times 480$ points in the first quadrant of Brillouine zone) at $T=300$\,K and the corresponding critical interaction scales were estimated from the maxima of the susceptibility in momentum space. 

\medskip

\paragraph{\textbf{Data availability.}}

All the data are available from the corresponding authors upon reasonable request.

\medskip

\paragraph{\textbf{Acknowledgments.}}
W.-H.K., Y.-T.H., M.-H.L. and D.K.~acknowledge National Science and Technology Council of 
Taiwan (grants No.~112-2112-M-006-019-MY3 and No.~114-2112-M-006-034-MY3) and National Center for 
High-performance Computing for computational resources.
Y.Z.~acknowledges support by the Swiss National Science Foundation (grant Nos. 224624).
J.G.~and~D.K. acknowledge support
from the EU Programme NextGenerationEU through the Recovery and Resilience Plan for Slovakia (project No.~09I03-03-V04-00682), from the programme IMPULZ 2021 funded by Slovak Academy of Sciences 
(Project IM-2021-26 SUPERSPIN), and by the Slovak Research and Development Agency 
(contract No.~APVV-24-0134).

\medskip

\paragraph{\textbf{Author contributions.}}
D.K. conceived the idea and supervised the project. W.-H.K. and Y.-T.H.,
under the supervision of M.-H.L. and D.K., performed numerical simulations of
spin filtering, magnetotransport, and persistent currents. Y.Z. and D.K.
calculated the altermagnetic spin susceptibilities and analyzed the resulting
instabilities. J.G. and D.K. developed the analytical theory of persistent
currents. D.K. provided the topological argument excluding genuine 2D open orbits in higher-order altermagnetic classes. W.-H.K., Y.-T.H., and D.K. drafted the main 
manuscript, with input from all authors. W.-H.K. and Y.-T.H., with assistance from 
J.G., prepared the Supplemental Material. All authors discussed the results and critically reviewed the manuscript.


\medskip

\paragraph{\textbf{Competing interests.}}
The authors declare no competing interests.


\medskip

\paragraph{\textbf{Corresponding authors.}}
Correspondence should be addressed to:
denis.kochan@phys.ncku.edu.tw, and, minghao.liu@phys.ncku.edu.tw.

\end{document}